\documentstyle[12pt,aasms,tighten]{article}
\newcommand{\cpl}{\kappa_{\alpha\beta\gamma}}
\newcommand{\cpla}{\kappa_{\alpha{\beta_1}{\beta_2}}}
\newcommand{\cplb}{\kappa_{\alpha{\beta}{\beta}}}
\newcommand{\BV}{Brunt-V\"ais\"al\"a~}

\begin{document}
% *************************************************************************
%                                 TITLE
% *************************************************************************

\title{\sc Nonlinear Damping of Oscillations in Tidal-Capture Binaries}

\author{Pawan Kumar\altaffilmark{1,2}}

\affil{Institute for Advanced Study, Princeton, NJ 08540}

\altaffiltext{1}{Permanent address:
        Massachusetts Institute of Technology,
	Department of Physics,
	Cambridge, MA~02139.
	Electronic mail: pk@brmha.mit.edu}
\altaffiltext{2}{Alfred P. Sloan Fellow \& NSF Young Investigator}

\author{Jeremy Goodman}

\affil{Princeton University Observatory,
	Peyton Hall,
	Princeton, NJ~08544 \\
	Electronic mail: jeremy@astro.princeton.edu}

% *************************************************************************
%				ABSTRACT
% *************************************************************************
\begin{abstract}

% Use the ``tighten'' option in documentstyle above to get single spacing:
%\baselineskip=14pt
We calculate the damping of quadrupole f- and low order
g-modes (primary modes) by nonlinear coupling to other modes of the star.
Primary modes destabilize high degree g-modes of half their frequency (daughter
modes) by 3-mode coupling in radiative zones.
For sunlike stars,
the growth time $\equiv\eta^{-1}\approx 4 E_{0,42}^{-1/2}$ days,
where $E_{0,42}$ is the initial energy of the primary mode in units
of $10^{42}~$erg, and the number of daughter modes
$N\sim 10^{10}E_{0,42}^{5/4}$.
The growth rate is approximately equal to the angular frequency of
the primary mode times its dimensionless radial amplitude,
$\delta R/R_*\approx 0.002E_{0,42}^{1/2}$.
Although the daughter modes are limited
by their own nonlinearities, collectively
they absorb most of the primary mode's energy
after a time $\sim 10\eta^{-1}$ provided $E_{0}> 10^{40}~\mbox{erg}$.
This is orders of magnitude smaller than usual radiative damping time.
In fact nonlinear mode interaction
may be the dominant damping process if $E_0\gtrsim 10^{37}~\mbox{erg}$.
These results have obvious application to
tidally captured main sequence globular cluster stars of mass
$\ge 0.5 M_{\sun}$; the tidal energy is dissipated in the radiative core
of the star in about a month, which is less than the initial orbital period.

Nonlinear mode coupling is a less efficient damping process for
fully convective stars, which lack g-modes. In convective stars most of
the tidal energy is in the quadrupole f-modes which nonresonantly excite
high order p-modes of degree 0, 2, and 4. The resultant short wavelength
waves are more efficiently dissipated. The nonlinear damping time for
f-modes is shown to be proportional to $1/E_0$; this damping time is
about 30 days for $E_0\approx 10^{45}~\mbox{erg}$
expected in tidal captures.
However, at such a large energy the system is very nonlinear: four-mode
and higher-order couplings are as important as three-mode couplings.
\end{abstract}

% *************************************************************************
%                               TEXT BODY
% *************************************************************************
\section{Introduction}
\baselineskip=14pt
Tidal interactions in highly eccentric close binary systems primarily excite
low order quadrupole gravity modes (g-modes) and the fundamental mode
(f-mode),
because these modes have periods comparable to the periastron passage time
and their eigenfunctions have the largest overlap with the tidal forcing
field [\cite{KAQ:95}].
The dissipation time scale for these modes, for main sequence stars,
are $10^4$ years or greater. A long damping time typically leads to a larger
mode amplitude for modes that are subjected to repeated tidal forcing, but
perhaps more interestingly offers the possibility that in tidal interactions
energy can be transferred from the modes to the orbit as well as from the
orbit to the modes, causing a chaotic evolution of the orbit [\cite{M:95}]
in some cases. In the case of tidal capture binaries, this process can lead to
ionization of the system in subsequent encounters
[\cite{K:92}, \cite{M:95}.]

Given the large amplitudes of modes in close binary systems, one
expects nonlinear interactions amongst modes carrying most of
the energy and others, which are tidally not excited, to be
an important process of draining energy from the primary modes.
We investigate this process and assess its significance in close binary
systems.

The next section describes the basic equations for mode couplings and
discusses the most significant nonlinear process for energy transfer
from modes of a radiative star. In section \S3 we discuss the
nonlinear mode couplings for a convective star.
Applications to tidal capture are discussed in \S4, and our
main results are summarized in \S5.
The details of the calculation of nonlinear coupling
coefficients are presented in Appendix A.

% *********** SECTION 2 *****************
\section{Nonlinear Interactions: Stars with Radiative Zones}

The leading-order nonlinear interactions of stellar modes are
3-mode couplings.
These can be derived by expanding the Lagrangian density for
adiabatic motions in powers of the fluid displacement.
If the expansion is truncated at second order, the standard
Euler-Lagrange procedure yields linear equations of motion,
which one solves for the normal modes.
When the third-order parts of the Lagrangian density are retained,
the equations of motion for the mode amplitudes contain
quadratic terms:
\begin{equation}
\ddot A_\alpha + \omega_\alpha^2 A_\alpha = -\omega_\alpha^2
    \sum_{\beta,\gamma} \cpl A_\beta A_\gamma,
\label{eq:gen}
\end{equation}
where $\omega_\alpha$ \& $A_\alpha$ are mode frequency and amplitude.
The 3-mode coupling coefficient, $\cpl$, is completely symmetric in
the three indices. It is obtained by substituting the normal mode
expansion of fluid displacement into the cubic Lagrangian density,
and integrating over the stellar interior (see Appendix A).
Since our eigenfunctions are normalized to unit energy, the
energy in mode $\alpha$ is $A_\alpha^2$. Note that equation (\ref{eq:gen})
with its right side set to zero describes free normal modes of oscillation.
We could represent tidal forcing by adding a term independent of the
amplitudes, but in this paper we are concerned with the
evolution of the modes after forcing has ceased.

We assume that initially all the pulsation energy resides in a few low order
modes. If the star has a radiative core, then these modes couple to high
degree g-modes and transfer energy away by parametric resonance instability.
This is discussed in the next sub-section. Even for completely convective
stars nonlinear energy transfer, involving non-resonant 3-mode interaction
with high frequency p-modes, can be an important process and is discussed
in \S3.

\subsection{Parametric instability}

Parametric instability arises when a mode is nonlinearly coupled to another
mode of approximately twice the frequency, or more generally when two low
frequency modes of infinitesimal amplitude
couple to a higher frequency mode of finite amplitude.
To analyze
this case we consider a single triplet interaction involving modes
$\alpha$, $\beta_1$ and $\beta_2$. Expressing mode amplitudes in the
following complex notation
\begin{equation}
 A_\alpha = A\exp(i\omega_\alpha t) + {\rm c.c.}, \quad\; A_{\beta_1} =
   B_1\exp(i\omega_{\beta_1}t) + {\rm c.c.}, \quad\;
   A_{\beta_2} = B_2\exp(i\omega_{\beta_2}t) + {\rm c.c.},
\end{equation}
and substituting this into eq.~(\ref{eq:gen}) we obtain the following
simplified equations
\begin{eqnarray}
\dot B_1 +\gamma B_1/2 & = & i 3\omega_{\beta_1}\cpla A B_2^*
       \exp(i\Delta\omega t), \\
       \dot B_2 + \gamma B_2/2 & = & i 3\omega_{\beta_2}\cpla A B_1^*
       \exp(i\Delta\omega t),
\end{eqnarray}
where $\Delta\omega\equiv\omega_\alpha-\omega_{\beta_1}-\omega_{\beta_2}$,
$\gamma$ is the linear dissipation rate of these modes,
and the factor of 3 appears for combinatorial reasons.
Rapidly oscillating terms as well as second
time derivative of mode amplitudes were discarded in deriving these equations.
At initial times when the amplitudes of modes $\beta_1$
and $\beta_2$ are infinitesimal, we can take the amplitude of mode $\alpha$
to be fixed ($A_0)$, and solve the above two equations for the growth rate
($\eta$) of mode amplitudes, which is
\begin{equation}
\label{eq:growth}
\eta = \frac{1}{2} \left[ \sqrt{9 A_0^2\cpla^2 \omega_{\beta_1}
       \omega_{\beta_2} - (\Delta\omega)^2} - \gamma \right].
\end{equation}
The coupling coefficient $\cpla$ is largest when the modes $\beta_1$
and $\beta_2$ have nearly the same radial order $n$ and spherical-harmonic
degree $\ell$, whence
$\omega_{\beta_1}\approx\omega_{\beta_2}\approx\omega_\alpha/2$.
We see that parametric instability sets in provided that
\begin{equation}
\label{eq:instrip}
\bigl(\Delta\omega\bigr)^2 < 9\cpla^2 A_0^2 \omega_{\beta_1}\omega_{\beta_2}
   -\gamma^2,
\end{equation}
Ignoring the damping for now, we obtain
a very simple criterion for parametric instability:
\begin{equation}
\label{eq:resonance}
\bigl|\Delta\omega\bigr| < 3 \omega_\alpha A_0 \cpla/2.
\end{equation}

The product $A_0 \cpla$ is a dimensionless quantity which is a measure of
the nonlinearity of the system.
The coupling coefficients are calculated using equation (\ref{eq:couplings})
of Appendix A, which is a somewhat long expression.
However, simple scaling argument shows
that $\kappa_{\alpha\beta\beta}\sim G^{-1/2}R_*^{1/2}M_*^{-1}$ for a low
order quadrupole mode coupled to two high order g-modes (where
$M_*$ and $R_*$ are stellar mass and radius), so that
$\cpla\approx 10^{-24}~\mbox{erg}^{-1/2}$ for a solar-mass main sequence star.
This is in good agreement with detailed numerical calculations (Table 1).
If the energy $E_0=A_0^2$ in the f-mode is $10^{42}~\mbox{erg}$
then parametric
resonance is possible provided that we can find a triplet with
$\Delta\omega$ less than $0.15\%$ of the frequency of the f-mode.
We discuss below the possibility of finding such resonant triplets
involving high degree g-modes.

To  put this energy in perspective, note that
the dimensionless radial amplitude of the f-mode is
$\delta R/R_*\approx 2\times 10^{-3} E_{0,42}^{1/2}$, where
$E_{0,42}\equiv E_0/(10^{42}\mbox{erg})$.
Tidal capture is expected to result in motions
$\delta R/R_*\sim 0.1$,  $E_0\sim 10^{45}~\mbox{erg}$.
For comparison, the most energetic five-minute oscillations
of the Sun have about $10^{28}~\mbox{erg}$, and the total
energy in all solar modes is $\approx 10^{33}~\mbox{erg}$ [\cite{GMK}].
It also follows from these scalings that
$\Delta\omega/\omega\approx 2\eta/\omega\approx\delta R/R_*$.

The dispersion relation for g-modes is [e.g. \cite{CDB:91}]
\begin{equation}
\label{eq:pnl}
P_{n\ell} = \frac{P_0}{2\sqrt{\ell(\ell+1)}} \bigl( 2 n + \ell - \delta
  \bigr) + \frac{P_0^2}{P_{n\ell}}\left[ V_1 + \frac{V_2}{\ell(\ell+1)}\right],
\end{equation}
where
\begin{equation}
P_0 = \frac{2\pi^2}{\int_0^{r_1} dr\, N_B/r},
\end{equation}
$P_{n\ell}$ ($=2\pi/\omega_{n\ell}=1/\nu_{n\ell}$)
is the mode period, $N_B$ is the \BV frequency, $r_1$
is the upper turning point of the mode,
$\delta\approx 5/6$, and $V_1$ and $V_2$ are constants of order 0.5 and 6
respectively for $1~M_{\sun}$ main sequence star. Note that mode frequency
goes to zero as $n/\ell$ goes to infinity, and the frequency approaches
$P_0^{-1}$ as $n/\ell$ tends to zero. Furthermore, it follows from
equation (\ref{eq:pnl})
that the frequencies of modes of large degree are function
of $n/\ell$, and that the frequency spacing between modes of adjacent $n$
for a fixed $\ell$, $\delta\nu_{n\ell}$, is approximately equal to
$\nu_{n\ell}^2 P_0/\ell$.
The total number of distinct frequencies $\nu_{n'\ell'}$ for which both
$n'$ and $\ell'$ are of order $\ell$ scales as $\ell^2$.
These frequencies are distributed between $0$ and $N_B$.
Statistically therefore, given a typical primary frequency $\nu_\alpha<2N_B$,
the lowest $\ell$ such that $|\nu_{n\ell}-(\nu_\alpha/2)|<\delta\nu$
should be $\ell\sim \ell_{\min}\sim\sqrt{\nu_\alpha/2\delta\nu}$.
It is worth noting that the frequency of the $\ell=2$ f-mode (strongly
excited during tidal capture) is in fact less than $2N_B$ in Sun-like
stars, so resonant g-modes will exist.

The radiative damping time of long-wavelength g-modes is of the order
of the photon diffusion time or thermal time $t_{th}$, which for a solar
type star is about $10^7$ years, and it decreases as $\ell^{-2}$ with
increasing mode degree. (The mode damping time for a completely radiative
star decreases much more rapidly with $\ell$ since the outer turning
point of g-modes moves outward with increasing degree. However, main sequence
globular cluster stars have a convective envelope of thickness $\gtrsim$
10\% of the stellar radius, and for these stars the radius of the outer
turning point of high-degree g-modes is independent of $\ell$ at fixed
frequency, and the damping time scales as $\ell^{-2}$). Thus the damping
rate of a mode of degree $\ell$ (and $n\sim\ell$) is
\begin{equation}
\label{eq:gamrad}
\gamma_\ell \sim \frac{\ell^2}{t_{th}}.
\end{equation}
Parametric instability can exist only if there are modes that have both
$\gamma_\ell$ and $|\Delta\omega|$ less than $3\omega_\alpha A_0 \cpla/2$
[see eq.~(\ref{eq:growth})].
This dual condition places a lower limit to the value of $A_0 \cpla$
at which parametric instability occurs, which we now calculate.
By the arguments of the last paragraph, the
minimum degree at which the resonance condition (\ref{eq:resonance})
is met is
\begin{equation}
\label{eq:lmin}
\ell_{\min} \approx \left( \frac{1}{A_0\cpla}\right)^{1/2}.
\end{equation}
The maximum degree is set by the requirement that mode damping be
less than $3\omega_\alpha A_0 \cpla/2$, which gives
\begin{equation}
\label{eq:lmax}
\ell_{\max} \approx \left( \frac{3 t_{th}\omega_\alpha A_0\cpla}{2}
   \right)^{1/2}.
\end{equation}
Obviously $\ell_{\max}\ge\ell_{\min}$ in order that there exist g-modes
resonantly coupled to the primary mode. This leads to the following
condition on mode amplitude for parametric instability to operate:
\begin{equation}
\label{eq:parains}
A_0 \cpla \gtrsim \sqrt{ \frac{1}{3\omega_\alpha t_{th}}}
\end{equation}
For a solar-type main sequence star, typically
$\cpla\sim10^{-24}~{\rm erg~s}^{-1/2}$, $t_{th}\sim 10^{14}~\mbox{s}$,
and $\omega_\alpha\approx 2\times10^{-3}~\mbox{rad/s}$.
Therefore, the f-mode must have energy
$\gtrsim 10^{36}~\mbox{erg}$ in order to
spawn a lower frequency mode via parametric instability.
Solar p and f modes lie at least seven orders of magnitude below
this threshold (see above).

So far we have neglected stellar rotation. Rotation lifts the
degeneracy of mode frequencies with respect to the spherical-harmonic
order $m$: thus we have $\nu_{n\ell m}$ instead of $\nu_{n\ell}$.
This enlarges the number of modes available
for resonant non-linear interactions, including parametric instability.
For a star rotating rigidly with angular velocity $\Omega_0$
the frequency splitting for high degree g-modes in the rotating frame
of the star, $\delta\nu_{\rm rot}\equiv \nu_{n,\ell,\ell}-\nu_{n,\ell,-\ell}$,
is approximately $\Omega_0/2\pi\ell$, whereas if the mean differential
rotation in the radiative interior of the star is $\overline{\Delta\Omega}$,
then $\delta\nu_{\rm rot}\sim\ell\overline{\Delta\Omega}/2\pi$
(see Appendix B).
Thus for a rigidly rotating star rotational splitting becomes important
at mode degree $\ell$ such that $\delta\nu_{\rm rot}\approx
\nu/\ell^2$, or $\ell \approx 2\pi\nu/\Omega_0\approx 10^3$ for a star like
the Sun. For $\ell$ greater than this value one can find modes with frequency
within $\Omega_0/\ell^2$ of some desired value, which is much better
than what we get for a nonrotating star.
Note that even a very small differential
rotation in radiative interior leads to a large value of $\delta\nu_{\rm rot}$.
However, in the case of the Sun, where the internal rotation is known
from measurements of p-mode frequency splitting, the
differential rotation in the radiative interior appears to be close to
zero. Nevertheless, in rotating stars one can find modes
with smaller values of $|\Delta\omega|$ and so
parametric instability persists down to smaller mode energies.

\subsection{Energy transfer to small-scale modes}

The astrophysical importance of the parametric instability depends upon
the amplitude at which nonlinearities intervene to stop exponential growth.
In simple mechanical systems, it is common for parametric instabilities
to saturate because the natural frequency of the daughter mode is
a function of amplitude.
Although the same mechanism applies in principle to g modes, it
does not affect the linear growth rate until the energy in a
single daughter mode is comparable to that in the primary mode, as
discussed in more detail in Appendix C.
Other nonlinearities become important at much smaller amplitudes.

As the energy of the primary mode decreases to balance the increasing
energy in all of its parametrically unstable daughters,
equations (\ref{eq:growth}) and
(\ref{eq:instrip}) remain true if we replace the initial amplitude ($A_0$)
with its current value ($A_\alpha$).
Hence both the width of the unstable resonance and the peak growth rate
decrease with $A_\alpha$.
Initially unstable daughters become stable.
It can be shown that parametric resonance involving a single daughter mode,
where $\beta_1=\beta_2$,
leads to periodic transfer of energy back and forth between the daughter
and the primary.

We now discuss the generalization of this result to the more realistic
case where a large number of modes ($\beta$) of angular frequency
$\omega_\beta\approx\omega_\alpha/2$ are coupled to the mode $\alpha$
of frequency $\omega_\alpha$.
At present, however, we ignore direct interactions among
the daughters themselves, as well as parametric decay of the
daughters into ``granddaughters.''
For clarity we write down the equations for
modes $\alpha$ and $\beta$ below:
\begin{equation}
\label{eq:modelA}
\dot A_\alpha = i\frac{3}{2}\omega_\alpha\sum_\beta
\kappa_{\alpha\beta\beta} B_\beta^2
   \exp(-i\Delta\omega_\beta t),
\end{equation}

\begin{equation}
\label{eq:modelB}
\dot B_\beta + \gamma_\beta B_\beta = i 3\omega_{\beta}
\kappa_{\alpha\beta\beta} A_\alpha  B_\beta^*
    \exp(i\Delta\omega_\beta t)
\end{equation}
where
\begin{equation}
\Delta\omega_\beta \equiv \omega_\alpha - 2\omega_\beta.
\end{equation}

It is straightforward to show from these equations that in the absence of
damping the sum of the energy in all modes is conserved,
i.e.\footnote{ \baselineskip=10pt Strictly speaking, the conserved energy
includes terms
of cubic and higher order in the amplitudes. In a hamiltonian formalism,
the latter terms are responsible for mode coupling
and for transfer of energy among resonant linear modes.
For a weakly nonlinear system, however, terms beyond the quadratic
can be neglected when evaluating the total energy.}
\begin{equation}
|A_\alpha|^2 + \sum_\beta |B_\beta|^2 = A_0^2 \equiv E_0.
\end{equation}
Furthermore, it can be shown that nonlinear interactions lead
eventually to equipartition of energy among all the coupled modes.
Thus the steady state energy of modes is
\begin{equation}
\label{eq:equipart}
\langle E_\alpha \rangle = \langle E_\beta \rangle \approx E_0/N,
\end{equation}
where $N$ is the number of modes involved in parametric resonance with
mode $\alpha$.
Initially, when all the energy is in mode
$\alpha$, daughter modes grow exponentially, and the characteristic time
for the energy in mode $\alpha$ to halve is of order $\eta^{-1}$
[eq.~(\ref{eq:growth})].
We shall call this energy-halving timescale $t_{\rm nl}$, since it derives
from nonlinear mode couplings.

In more detail, $t_{\rm nl}$ depends logarithmically on the mean
initial energy per daughter mode, $E_{\beta,0}$.
The latter energy has at least the thermal
value $kT\sim 10^{-9}~\mbox{erg}$, whence
$t_{\rm nl}<\ln(E_0/NkT)/2\eta\approx 50/\eta$.
Since the g-modes may be stochastically excited by macroscopic
processes such as semiconvection in the core or convection in the
envelope of the star, it is probably more realistic to assume
$t_{\rm nl}\sim 10/\eta$, or approximately $50 E_{\alpha,42}^{-1/2}~$days.

In the simplified model of eqs.~(\ref{eq:modelA})-(\ref{eq:modelB}),
complete equipartition [eq.~(\ref{eq:equipart})] is obtained only on a
timescale $\gg t_{\rm nl}$, because the nonlinear interaction
rate decreases in proportion to the primary amplitude; furthermore,
phase coherence between the primary and the daughters tends to be lost.
Because of damping and because of coupling to granddaughter modes,
equipartition may never be achieved in a real star.

We have verified that our model behaves as described
by numerically integrating the differential equations
(\ref{eq:modelA})-(\ref{eq:modelB}),
and the results are shown in fig. 1.

We can compare the nonlinear timescale to the timescale for decay
of the primary mode by linear radiative damping.
The former is
\begin{equation}
\label{eq:tnl}
t_{\rm nl} \approx {10\over\eta} \approx {40\over 3 \omega_\alpha A_0 \cpla}
\approx {2\over \nu_\alpha A_0 \cpla}.
\end{equation}
If the radiative damping time for the primary mode is $1/\gamma_\alpha$,
then parametric instability is the dominant damping process for the mode
provided that
\begin{equation}
A_0 \cpla \gtrsim \frac{2\gamma_\alpha}{\nu_\alpha}.
\end{equation}
The damping time for solar f and low order g modes is about $10^4$ years.
By this comparison, nonlinear damping would appear to be more rapid than
linear damping if the energy of the primary is greater than
$10^{34}~\mbox{erg}$, which is less than the minimum energy at which
parametric instability can exist [eq.~\ref{eq:parains})]. So the nonlinear
process should always dominate whenever it is allowed at all.
But we shall revise this conclusion below.

We now consider the long-term evolution of the system. After a period
$\sim t_{\rm nl}$, the energy in the
primary mode drops by a factor $\gtrsim 2$ and is shared among
$N\sim\ell_{\max}^3(\Delta\omega/\omega)\sim\ell_{\max}^3/\ell_{\min}^2$
parametrically resonant daughter modes.
Using eqs.~(\ref{eq:lmin})-(\ref{eq:lmax}) and the typical values of
$\cpla$, $t_{th}$, and $\omega_\alpha$ cited above, we estimate
$N\sim 10^{10}E_{0,42}^{5/4}$.
The time scale for further
energy decline is set by the daughter modes, which both dissipate energy
by radiative process and also spawn lower frequency modes.
The radiative damping time for g-modes of $\ell\sim 10^{4}$ is on order
a week; and in fact this time is $\sim\eta^{-1}$
for modes of $\ell\sim\ell_{\max}$.
Energy dissipated this way can clearly never return to global oscillations
of the star.

Additionally, energy can also
be removed from daughter modes if they themselves parametrically
destabilize modes of still lower frequency---``granddaughter modes''.
The parametric growth rate of a granddaughter mode ($\gamma$)
due to nonlinearity in the daughter mode ($\beta$)
is $\eta'\approx \omega_\beta\kappa_{\beta\gamma\gamma}A_\beta$.
Energy transfer to the granddaughters will significantly retard
the growth of the daughter modes when $\eta'\ge\eta$,
or in other words $\kappa_{\beta\gamma\gamma}A_\beta\ge
\kappa_{\alpha\beta\beta}A_0$.
Now it follows from an analysis of the 3-mode couplings
that $\kappa_{\beta\gamma\gamma}\sim(\ell_\beta/\ell_\alpha)
\kappa_{\alpha\beta\beta}$ if $\ell_\beta\gg\ell_\alpha\approx 2$
(Appendix A).
Hence secondary parametric instabilities become important when
$A_\beta\sim A_0/\ell_\beta$.
Most of the daughter modes have degree $\ell_\beta\sim\ell_{\max}$, and after
a time $\sim t_{\rm nl}$ each has a typical amplitude
$A_\beta\sim N^{-1/2}A_0\sim(\ell_{\min}^2/\ell_{\max}^3)^{1/2}A_0$.
This is smaller than $A_0/\ell_{\max}$ if $\ell_{\max}>\ell_{\min}^2$.
 From eqs.~(\ref{eq:lmin})-(\ref{eq:lmax})
we find $\ell_{\max}/\ell_{\min}^2\approx 25 E_{0,42}^{3/4}$.

Thus in the regime $E_0\ll 10^{40}~$erg, most of the
daughter modes will be limited to energies much less than the
equipartition value ($E_0/N$) by secondary parametric instabilities.
We can estimate how much this process should increase the nonlinear
damping time of the primary mode.
We assume that each daughter mode ($\beta$)
continues to absorb energy from the primary
at a rate $\approx 2\eta E_\beta$ even after $E_\beta$ has saturated.
If saturation occurs at $E_\beta\sim (A_0/\ell_\beta)^2$,
the primary loses energy at the rate
\begin{equation}
\label{eq:netdamp}
\dot E_\alpha\sim -\eta
\int\limits_{\ell_{\min}}^{\ell_{\max}}\frac{E_\alpha}{\ell_\beta^2}
\left(\frac{\ell_\beta}{\ell_{\min}}\right)^2 d\ell_\beta
\sim -\eta E_\alpha\left(\frac{\ell_{\rm max}}{\ell_{\min}^2}\right),
\end{equation}
instead of $\dot E_\alpha=-\eta E_\alpha$ as for
$E_\alpha\gg 10^{40}~\mbox{erg}$.
The modified nonlinear damping rate implied by eq.~(\ref{eq:netdamp})
scales as $E_\alpha^{5/4}$ instead of $E_\alpha^{1/2}$.
Comparing this modified damping rate to the linear radiative damping
of the primary mode ($\sim 10^4~\mbox{yr}$), we find that
nonlinear damping  continues to dominate provided
$E_\alpha\gtrsim 10^{37}~\mbox{erg}$, which is only an order of magnitude
above the threshold for parametric instability (\S 2.1).

At the energies expected of tidal capture
($E_0\sim 10^{42-45}~\mbox{erg}$),
the average energy per daughter mode is so small, because their number
$N\propto E_0^{5/4}$ is so large, that the growth
rate of the granddaughter modes can be neglected for the most part.
However, secondary instabilities will prevent the
fastest-growing daughters---those near the center of the instability
strip (\ref{eq:instrip})---from exceeding an amplitude $\sim A_0/\ell_\beta$.
By contrast, in the simplified model of
eqs.~(\ref{eq:modelA})-(\ref{eq:modelB}) where secondary
instabilities are ignored, the single fastest-growing mode briefly
attains almost the entire energy of the system.

Low order g-modes of a radiative star carry a substantial amount of energy
as a result of tidal interaction. The parametric
instability described above will operate on these modes just as well,
and independently of the f-modes, causing a loss of their energy on a time
scale similar to the f-mode. This is discussed in more detail in \S4.

\section{Nonlinear energy dissipation in convective stars}

The parametric nonlinear interaction mechanism described in the last
section does not apply to fully
convective stars, which do not have stable g-modes.
However, even for these stars nonlinear mode coupling can
be an important dissipation mechanism for the f-mode energy.
The energy in f-modes of a convective star after tidal capture is often
an order of magnitude larger compared to a radiative star of same mass and
orbital parameters, and thus the nonlinear parameter, $A_\alpha\cpl$,
is larger for convective stars. In this case even the
normally very inefficient process of nonlinear coupling of a f-mode to
higher frequency p-modes can transfer energy to short wavelength waves where
it is dissipated on a relatively short time scale.
The nonlinear response to the f-mode can be treated as
a forced excitation of all p-modes of degree $\ell=$ 0, 2, and 4,
in which the forcing function is quadratic in the f-mode amplitude
and oscillates at twice the f-mode frequency.
We calculate this response by solving an inhomogenous wave equation
for the sum of the p-mode amplitudes with the nonlinear forcing function
as the inhomogenous term.
The latter can be obtained either from the
Lagrangian given in Appendix A, or by expanding mass, momentum and energy
equations to second order in wave quantities [e.g. \cite{Dz:82}].
The solution of the inhomogeneous
equation is shown in figure 2 together with the f-mode eigenfunction.
The turbulent viscous, and thermal, dissipation of
wave energy is calculated using this wavefunction.
Note the generation of significant power at short wavelengths and the
relatively large displacement amplitude and its derivative in the outer
part of the star; these effects enhance the dissipation of f-mode energy.
The resulting nonlinear dissipation time for the
f-mode by coupling to p-modes is
\begin{equation}
\label{eq:tffp}
t_{ffp} = 3\times10^4\, E_{42}^{-1}\;\; \mbox{days},
\end{equation}
where $E_{42}$ is f-mode energy in units of $10^{42}\mbox{erg}$.

We can estimate the damping rate in another way that is less
accurate but provides some insight.
In the present special case of the general 3-mode interaction (\ref{eq:gen}),
we drop all nonlinear terms except the square of the f-mode amplitude.
The result is
\begin{equation}
\ddot A_\beta + \omega_\beta^2 A_\beta = -3\omega_\beta^2
    \kappa_{\beta\alpha\alpha} A_\alpha^2,
\end{equation}
where $A_\alpha$ is the amplitude of the f-mode which we take to be
$A_0\sin(\omega_\alpha t)$. Substituting this into the above equation we
find for the amplitude of mode $\beta$,
\begin{equation}
A_\beta = -\frac{3\omega_\beta^2 A_0^2 \kappa_{\beta\alpha\alpha}
   \cos(2\omega_\alpha t) }{ 2(\omega_\beta^2 - 4\omega_\alpha^2)}.
\end{equation}
The net displacement associated with the p-mode amplitudes $A_\beta$ is
\begin{equation}
\vec\xi = \sum_\beta A_\beta \vec\xi_\beta,
\end{equation}
which can be used to calculate the dampings. The function $A_\beta
\vec\xi_\beta$ has alternating signs for even and odd order modes near the
surface of the star where most of the damping is taking place. This can
lead to significant cancellation in the total displacement.
Nevertheless we proceed by
assuming that different modes can be added in quadrature.
The energy in mode $\beta$ is
\begin{equation}
\langle E_\beta \rangle = \frac{9}{16}\frac{\omega_\beta^4 (A_0
   \kappa_{\beta\alpha\alpha})^2 A_0^2}{(\omega_\beta^2 -
   4\omega_\alpha^2)^2}
   \left[ 1 + \frac{4\omega_\alpha^2}{\omega_\beta^2}\right].
\end{equation}
If the rate of dissipation of energy of a mode $\beta$ driven
at frequency $2\omega_\alpha$ is $\Gamma_\beta$, then the rate of
energy dissipation of the f-mode (mode $\alpha$) can be calculated
from this last equation, and the resulting dissipation time is
\begin{equation}
\label{eq:tffpincoh}
t_{ffp} \sim\left(A_0^2 \sum_\beta \Gamma_\beta
   \kappa_{\beta\alpha\alpha}^2\right)^{-1}.
\end{equation}
The 3-mode coupling coefficients, $\kappa_{ffp}$, involving a quadrupole f
and a radial p mode for a solar model, have a typical value
$10^{-24}~\mbox{erg}^{-1/2}$, and the
damping time for high order p-modes (1/$\Gamma_p$) is about 10 days.
Since there are about 100 p-modes of degree 0, 2, and 4 coupled to the
quadrupole f-mode, we see from equation (\ref{eq:tffpincoh})
that for a f-mode of energy $10^{42}~\mbox{erg}$, the non-linear
damping time is about $10^4$ days which is within a factor of a few of the
numerically calculated value given in eq.~(\ref{eq:tffp}). For comparison
we note that the linear turbulent viscous damping time of f-mode is about
$10^4$ years.

The energy in the f-mode of a 0.5 solar
mass polytropic star of index 1.5, tidally captured by a one solar
mass star, is about $10^{45}$ erg ($\delta R/R_*\approx 0.1$),
and so its nonlinear dissipation [\ref{eq:tffp})]
time is about 30 days. At such high energies, however, the system is
very nonlinear: that is, four mode couplings are as important as the
three mode couplings.
%(the 4-mode coupling coefficient $\kappa_{fffp}\sim$ $10^{-46}$ erg$^{-1}$,
%and so the dimensionless number $A_0 \kappa_{fffp}/\kappa_{ffp}$ which is a
%measure of the relative importance of 4-mode to 3-mode couplings is
%of order 1).
Figure 2 shows that for this energy the amplitude of the p-mode response is
about the same (or greater) as the f-mode amplitude throughout the star,
and thus higher order couplings must be equally important.
In this case
it is very likely that high order p-modes are resonantly excited leading
to even more efficient dissipation to f-mode energy. Moreover, since most
of the energy is dissipated in the outermost envelope of the star at a rate
of $\sim 10^{39}~\mbox{erg~s}^{-1}$,
which is more than can be transported by convection,
it will cause the envelope to expand away from the star. This is a
further indication of the breakdown of the weak nonlinear calculation
presented here.

In addition to the nonlinear effect arising from the third order fluid
Lagrangian, considered here, there is also
nonlinearity associated with the thermal dissipation process itself. It can
be shown that both of these are of the same order, and thus a neglect of
the latter does not change our result by more than a factor of a few,
which is within the uncertainty of the calculation of this section.

\section{Application to capture binaries in globular clusters}

Main sequence stars of mass greater than 0.5 $M_{\sun}$ have a sizable
radiative core (see table 2) that
supports a dense family of g-modes. When such stars are tidally captured,
resulting in a highly eccentric orbit with separation between the stars
at periastron of only a few stellar radii initially, their quadrupole f and
low order g modes are excited to large amplitudes. The linear radiative
damping time for these modes is of order $10^4$ years. However these modes
are resonantly coupled to high degree g-modes by parametric instability
which provides a very efficient way of dissipating their energy (see \S2).
We now estimate the time for the energy of the low order modes to
decay due to nonlinear interaction using the results of \S2.

The main ingradient required for nonlinear dissipation calculation is the
3-mode coupling coefficients, $\cpla$, which are computed using
equation (\ref{eq:couplings}). The numerical
calculation of $\cpla$ is carried out for a simplified polytropic
stellar model, which is widely used in tidal capture and evolution
calculations, as well as a standard solar model. The main sequence
globular cluster stars of mass greater than about 0.5 $M_{\sun}$ have a
radiative core (see table 2) and can be crudely modelled as a polytrope of
index 3 [cf. \cite{PDD:92}, \cite{LRS:94}, \cite{De:95}].
The coupling coefficients involving low degree g-modes
($\ell_\beta\le 40$) are calculated
using eigenfunctions obtained by solving the wave equation, whereas for
high degree g-modes we use the WKB solution which is an excellent
approximation.
We find that $\cplb$ for the Sun and 1.0 $M_{\sun}$ polytropic star of
index 3 agree to within a factor of two for $\ell_\beta\lesssim 100$
(the polytropic model yields a higher value).
However, coupling coefficients increase with $\ell_\beta$ for
polytropic models but not in the case of the Sun or a realistic
stellar model. This is because g-modes are trapped in the radiative interior
of low mass main sequence stars,
whereas in a polytropic model, where the \BV frequency increases
monotonically with distance from the center, g-modes propagate closer and
closer to the surface with increasing mode degree.
One should therefore use a realistic stellar model for the purpose of
nonlinear mode interaction calculation. Since $\cpla\sim G^{-1/2}R_*^{1/2}
M_*^{-1}$ depends only weakly
on stellar mass we use a solar model to compute coupling coefficients
that are used for lower mass globular cluster stars; the true coupling
coefficients are larger by about 20\%. Table 1 shows the value of
$\kappa_{\alpha\beta\beta}$ for a selected number of modes. Note that
a typical value of coupling coefficient is $10^{-24}$ erg$^{-1/2}$ which is
in good agreement with the rough estimate of $G^{-1/2}R_*^{1/2}M_*^{-1}$.

The dependence of $\cplb$ on $m_\beta$ is entirely contained in
Wigner 3-J symbol [see eq. (\ref{eq:triplL})],
which for large $\ell_\beta$ scales as
$\cos(\pi m_\beta/2\ell_\beta)$. Thus a good fraction of modes of
all possible $m$, for a given $\ell$, are equally well coupled to
the f and low order g modes. Finally, we note that the coupling coefficient
decreases with increasing $\Delta n\equiv |n_{\beta_1}-n_{\beta_2}|$
and is neglegible if $\Delta n\gtrsim 10$.

The energy input rate per orbital period, for several different modes,
are shown in figure 3 as a function of distance between the stars at
periastron. This was calculated using a scheme similar to \cite{PT:77}.
In general most of the
energy is in some g-mode which has a period close to the periastron
passage time. Substituting mode amplitudes and coupling coefficients
into equation (\ref{eq:tnl}) we obtain nonlinear damping times which are
displayed in figure 4 for several modes as a function of periastron
separation. Note that for mode energy of $10^{42}$ erg (corresponding to
dimensionless surface displacement amplitude $\delta R/R_*\approx$
2x$10^{-3}$; see Table 1)
the nonlinear damping time is 50 days. During this time the mode energy
decreases by a large factor, which is shared equally by about $N\sim
10^{10}$ high degree g-modes that have been parametrically excited via
nonlinear mode coupling.
It follows from equation (\ref{eq:tnl}) that the nonlinear
damping time scales as $E_\alpha^{-1/2}$, and thus for mode energy of
$10^{44}$ erg the damping time is only 5 days.
For comparison we note that the radiative damping time
for low order quadrupole modes are of order $10^4$ years. Thus
nonlinear dissipation dominates even when periastron distance is $4R_*$
(see fig. 3 \& 4).
In fact the nonlinear dissipation rate is larger than the linear radiative
damping so long as mode energy is greater than about $10^{37}$ erg.
It should be emphasized that energy removed from a mode by nonlinear
interactions is distributed over a large number of g-modes. The energy is
eventually thermalized on the dissipation time scale of high degree g-modes,
which for modes of $\ell\sim\ell_{\max}$ is of the same order as $t_{\rm nl}$.
How the star adjusts to the energy deposited in its core on a short time
scale is outside of the scope of this work. However, we note that even
$10^{44}$ erg of energy deposited in the radiative core is a small fraction
of its thermal energy content, and thus will have a minor effect on the
structure of the star.

Main sequence globular cluster stars of mass less than about 0.5 $M_\odot$
are almost fully convective, and therefore the coupling of f-mode to g-modes
is very weak. The dominant nonlinear coupling of the f-mode in this case,
as described in \S3, is with p-modes. Physically this nonresonant
mode coupling results in power on smaller length scale, and also the value
of the gradient of the wave function generated can be considerably larger
than the original f-mode wave function in the outer part of the star
(see fig. 2). This leads to more efficient dissipation of wave energy.
As discussed in \S3 the f-mode dissipation time due to this process is
about 3x$10^4$ E$_{42}^{-1}$ days.

\bigskip
\centerline{\bf 5. Summary}
\medskip
Tidally excited modes of a main sequence star in a highly eccentric orbit,
with periastron distance of 4 stellar radii or less, have energy about
$10^{42}$ erg or greater, or fractional change in stellar radius
of 0.2\% or more. The linear radiative and turbulent dissipation times for
these modes are about $10^4$ years. We have calculated the time needed
for energy of tidally excited modes to decrease due to nonlinear interaction
with other modes in the star for two different cases that are described below.

One of the cases considered is that of a star with a radiative core -- for
instance main sequence globular cluster stars of mass greater than about 0.5
M$_{\sun}$. The most important mode coupling in these radiative stars
involves high degree g-modes which are parametrically excited
by the low order quadrupole modes. The time scale for energy transfer for
this process is $2/(\nu_\alpha\kappa_{\alpha\beta\beta} E_\alpha^{1/2})$,
where $\kappa_{\alpha\beta\beta}\sim G^{-1/2} M_*^{-1} R_*^{1/2}\sim
10^{-24}$erg$^{-1/2}$ is 3-mode coupling coefficient. Thus
for $E_\alpha\approx 10^{42}$erg, or $\delta R/R_*\approx$ 2x$10^{-3}$,
the nonlinear damping time is 50 days. This is smaller by a factor of $10^6$
compared to the nonlinear damping time estimated by Kochanek (1992), who
had not considered parametric coupling. The energy removed from the
primary mode initially resides in roughly $10^{10}$ high degree g-modes
and is later dissipated in the radiative core of the star on a time scale
of about a year. Since this energy is small compared to the thermal energy
content of the core, it is unlikely to have much effect on the structure of
the star in one orbital passage.

The other case we have considered is that of a fully convective star,
such as a main sequence star of mass less than about 0.5 M$_{\sun}$.
Most of the tidal energy in these stars, which do not have stable g-modes,
resides in quadrupole f-modes. The f-modes couple to high
order p-modes of degrees $\ell=$0, 2, and 4 generating short wavelength waves
that significantly enhance the energy dissipation rate. The energy
of short wavelength p-modes is dissipated near the stellar surface by
turbulent and radiative viscosity. The nonlinear energy
dissipation time of f-modes is estimated to be $3\times10^4$
($10^{42}\mbox{erg}$
/$E_f$) days or $2\times10^{-1}$ $(\delta R/R_*)^{-2}$ days,
which is shorter than
the linear dissipation time when the mode energy is greater than about
$10^{40}$ erg. And it is also smaller than the nonlinear damping time
estimate of Kochanek (1992) by a factor of about $10^3$. For mode energy of
$10^{45}$ erg, that which is expected in tidal captures, the nonlinear damping
time is about 30 days. However, for such a large energy, 4-mode
5-mode, and higher-order mode coupling are
also important; that is, the weak nonlinear approximation is not valid.

\bigskip
{\bf Acknowledgment:} PK is grateful to Piet Hut for suggesting this
problem, and he is indebted to John Bahcall for
his support and hospitality at the IAS where this work was carried out.
We thank Peter Goldreich, Piet Hut, Steve Lubow, John Papaloizou,
and J.E. Pringle for useful discussions---the last three in particular
for prompting us to add Appendix C.
JG acknowledges NASA support under grant NAG5-2796.

\newpage

\appendix
\section{Adiabatic g-mode couplings in spherical geometry}

\newcommand{\Lag}{{\cal L}_3}
\newcommand{\divxi}{\vec\nabla\cdot\vec\xi}
\newcommand{\divXi}{\vec\nabla\cdot\vec\Xi}
\newcommand{\dep}{\delta p}
\newcommand{\Lf}{{\Lambda_f}}
\newcommand{\La}{{\Lambda_1}}
\newcommand{\Lb}{{\Lambda_2}}
\newcommand{\cd}{\hat\nabla}
\newcommand{\hg}{\hat g}
\newcommand{\pp}{{dp\over dr}}
\newcommand{\ppp}{{d^2p\over dr^2}}
\newcommand{\PD}{{\partial}}

The Lagrangian for inviscid, compressible fluid with internal energy
$E(s,\rho)$ is
\begin{equation}
{\cal L} = \rho\left( {v^2\over 2} - E(s,\rho) - U({\bf x})\right),
\end{equation}
where $v$ is fluid velocity, $s$, $\rho$, and $U$ are entropy,
density, and gravitational potential.
The three mode coupling is described by the fluid Lagrangian expanded to
third order in displacement and is given below

\begin{eqnarray}
\Lag &=& -{\textstyle 1\over2}\xi^i\xi^j\dep_{;i;j}-\dep_{;i}\xi^i\divxi-
{\textstyle 1\over2} (\divxi)^2\dep \nonumber \\
&+&{\gamma(\gamma-2)\over6}p(\divxi)^3 -{\textstyle 1\over2}p_{;i}\xi^i
(\divxi)^2 -{\textstyle 1\over2}p_{;i;j}\xi^i\xi^j\divxi \nonumber \\
&-&{1\over6}p_{;i;j;k}\xi^i\xi^j\xi^k
-{1\over6}U_{;i;j;k}\,\rho\,\xi^i\xi^j\xi^k
-{\textstyle 1\over2}\rho\xi^i\xi^j\delta U_{;i;j}.
\label{eq:lagxi}
\end{eqnarray}
Here $p(r)$ is the equilibrium pressure profile, $\xi$ is the fluid
displacement from its equilibrium position, semicolon denotes
covariant derivative (see Misner, Thorne, and Wheeler 1973 for
the definition of covariant derivative), and $\delta p$ is
the Eulerian pressure perturbation to first order in $\xi$:
\begin{equation}
\delta p\equiv -\vec\xi\cdot\vec\nabla p -{\gamma p\over\rho}\divxi.
\end{equation}
\noindent Let $\xi\to\xi+\Xi$, where $\Xi$ is a large-scale fundamental
or g-mode and $\xi$ becomes the sum of all small-scale modes.
Then the dominant (i.e., leading order in $\ell$ of small-scale modes)
terms are
\begin{eqnarray}
\label{eq:lagexp}
\Lag &=& \Xi^j_{~;i}\xi^i\dep_{;j}-{\textstyle 1\over2}\xi^i\xi^j
\left(\delta P_{;i;j}
+\rho\delta U_{;i;j}\right) -\divxi\,\xi^i\delta P_{;i}
-{\textstyle 1\over2}(\divxi)^2\delta P \nonumber \\
&+&{\gamma(\gamma-2)\over2}\,p(\divxi)^2\divXi
-{\textstyle 1\over2}p_{;i}\Xi^i(\divxi)^2
-p_{;i}\xi^i\divxi\,\divXi \nonumber \\
&-&p_{;i;j}\left({\textstyle 1\over2}\xi^i\xi^j\divXi+\xi^i\Xi^j\divxi
\right)
-\frac{1}{2}(p_{;i;j;k}+\rho U_{;i;j;k})\xi^i\xi^j\Xi^k
\end{eqnarray}
Capital letters are used to distinguish quantities pertaining to the
large-scale mode: e.g. $\delta P$ and $\delta U$ are the first-order
eulerian pressure and potential perturbations associated with $\Xi$.
(However $U$ unadorned with $\delta$ is the equilibrium potential.)
The 3-mode coupling coefficient is the integral of $\Lag$ over the star,
when the mode eigenfunctions are normalized to unit energy.

Integration by parts has been used in deriving formulae (\ref{eq:lagxi})
and (\ref{eq:lagexp})
to remove as many derivatives as possible from the small-scale
displacements $\vec\xi$, except where these occur as divergences
$\divxi$.
In general $\xi^i_{~;j}\sim\xi/\lambda$, where $\lambda\ll R_*$ is lengthscale
associated with the displacement.
Because the small-scale g-modes are nearly incompressible and isobaric,
however, $\divxi\sim\xi/H$ and $\delta p/p\sim \xi\lambda/H^2$, where
$H\sim R_*$ is the pressure or density scale height.

The next step is to integrate $\Lag$ over solid angle
($d^2\Omega=\sin\theta d\theta d\phi$).
As usual, the covariant components of the
angular displacements can be written as partial derivatives of a scalar
potential
[here and henceforth, greek indices range over the angular coordinates
$\{\theta,\phi\}$, whereas roman ones range over $\{r,\theta,\phi\}$],
\begin{equation}
\xi_\alpha=\PD_\alpha\psi,~~~~\Xi_\alpha=\PD_\alpha\Psi,
\end{equation}
and the displacements can be decomposed into angular harmonics:
\begin{eqnarray}
\xi^r&=&\sum\limits_{\Lambda}\xi^r_\Lambda(t,r) Y_\Lambda(\theta,\phi),~~~~~~~
\psi =\sum\limits_{\Lambda}\psi_\Lambda(t,r) Y_\Lambda(\theta,\phi),\\
\Xi^r&=&\sum\limits_{\Lf}\Xi^r_\Lf(t,r) Y_\Lf(\theta,\phi),~~~
\Psi =\sum\limits_{\Lf}\Psi_\Lf(t,r) Y_\Lf(\theta,\phi).
\end{eqnarray}

We use the single symbol $\Lambda$ for the degree and order
$(\ell,m)$ of a spherical harmonic i.e., $Y_\Lambda\equiv
Y_{\ell m}(\theta,\phi)$, and define $\Lambda^2\equiv\ell(\ell+1)$.
The projection of a function $f(t,r,\theta,\phi)$ onto $Y_{\ell m}$ is written
$f_\Lambda$ and is understood to be a function of $(t,r)$
(except that $Y_\Lambda$ denotes $Y_{\ell m}(\theta,\phi)$):
\begin{equation}
f_\Lambda\equiv\int f(t,r,\theta,\phi)Y^*_\Lambda(\theta,\phi)d^2\Omega.
\end{equation}
Enclosing a term in angle brackets means to integrate it over solid angle;
thus the above could be written $f_\Lambda=\langle Y^*_\Lambda f\rangle$.
The angular integral of three spherical harmonics is related
to the Wigner 3j symbol (see Edmonds, 1960),
\begin{equation}
\pmatrix{%
\ell_1 & \ell_2 & \ell_3 \cr
 m_1   & m_2    & m_3    \cr}
\end{equation}
by
\begin{eqnarray}
\label{eq:triplL}
\langle\Lambda_1\Lambda_2\Lambda_3\rangle&\equiv&
\langle Y_{\Lambda_1}Y_{\Lambda_2}Y_{\Lambda_3}\rangle\nonumber\\
&=&\left[(2\ell_1+1)(2\ell_2+1)(2\ell_3+1)\over 4\pi\right]^{1/2}
\pmatrix{%
\ell_1 & \ell_2 & \ell_3 \cr
 m_1   & m_2    & m_3    \cr}
\pmatrix{%
\ell_1 & \ell_2 & \ell_3 \cr
 0     &   0    & 0      \cr}.
\end{eqnarray}

To calculate $\langle\Lag\rangle$ from eq.~(\ref{eq:lagexp}), one needs
the following identities:
\begin{equation}
\label{eq:ident1}
\left\langle Y_{\Lambda_3}\,\hg^{\alpha\beta}(\cd_\alpha Y_{\Lambda_1})
(\cd_\beta Y_{\Lambda_2})\right\rangle
={\Lambda_1^2+\Lambda_2^2-\Lambda_3^2
\over 2}\langle\Lambda_1\Lambda_2\Lambda_3\rangle,
\end{equation}

\begin{equation}
\left\langle \hg^{\mu\alpha}\hg^{\nu\beta}
(\cd_\mu\cd_\nu Y_{\Lambda_3})(\cd_\alpha Y_{\Lambda_1})
(\cd_\beta Y_{\Lambda_2})\right\rangle
={(\Lambda_3^2)^2-(\Lambda_1^2-\Lambda_2^2)^2\over 4}
\langle\Lambda_1\Lambda_2\Lambda_3\rangle.
\end{equation}
Here $\cd$ is the usual covariant derivative on the sphere; that is,
$\cd_\gamma\hg_{\alpha\beta}=0$, where $\hg_{\alpha\beta}$ is the 2D
metric
\begin{equation}
\hg_{\alpha\beta}dx^\alpha dx^\beta= d\theta^2+\sin^2\theta d\phi^2,
\end{equation}
and $\hg^{\alpha\beta}$ is its inverse.

To establish Eq. (\ref{eq:ident1}), for example, note that
\begin{equation}
\hg^{\alpha\beta}(\cd_\alpha Y_{\Lambda_1})
(\cd_\beta Y_{\Lambda_2})= {1\over2}\left[\cd^2(Y_{\Lambda_1}
Y_{\Lambda_2})-Y_{\Lambda_1}\cd^2Y_{\Lambda_2}
-(\cd^2Y_{\Lambda_1})Y_{\Lambda_2}\right],
\end{equation}
and that
\begin{equation}
\cd^2Y_{\Lambda}\equiv\hg^{\alpha\beta}\cd_\alpha\cd_\beta Y_{\Lambda}
=\Lambda^2 Y_{\Lambda},
\end{equation}
and use integration by parts.
The 2D covariant derivative is not simply the restriction of the 3D
covariant derivative to the sphere,
but is related to it by formulae such as
\begin{eqnarray}
f_{;\alpha}&=&\cd_\alpha f=\PD_\alpha f,\nonumber \\
 v^r_{~;\alpha}&=&\cd_\alpha v^r - r^{-1} v_\alpha,
\PD_\alpha v^r - r^{-1} v_\alpha,\nonumber \\
v_{\alpha;\beta} &=&\cd_{\beta}v_\alpha -r\hg_{\alpha\beta}v_r,
\end{eqnarray}
where $f$ and $v^i$ are scalar and vector fields in 3D.
To avoid confusion, we raise and lower indices using $g_{ij}$ only:
for example, $v_\alpha\equiv g_{\alpha i}v^i$
(not $\hg_{\alpha\beta}v^\beta$).

Using these rules we integrate the third order Lagrangian over angle which
leads to the following expression

\begin{eqnarray}
&&\int\Lag d^2\Omega =
 \sum\limits_{\La,\Lb,\Lf}\langle\La\Lb\Lf\rangle \;\left\{
\vphantom{\left[\ppp\right]}\right.\nonumber \\
&&{{(\Lf^2)^2-(\La^2-\Lb^2)^2\over 4r^4}}
\left[\Psi_\Lf\psi_\La\dep_\Lb
-{\textstyle 1\over2}\left(\delta P+\rho\delta U\right)_\Lf\psi_\La\psi_\Lb
\right]~~+\nonumber \\
&&{{\La^2+\Lb^2-\Lf^2\over2r^2}}
\left[r\Xi^r_\Lf\psi_\La\dep_\Lb
-{\textstyle 1\over2}\left(r\PD_r\delta P+\rho\,r\PD_r\delta U
+{1\over r}\pp\divXi\right.\right.\nonumber \\
&&\left.\left.\qquad\qquad\qquad\qquad
+{d\over dr}\left({1\over r}\pp\right)\Xi^r\right)_\Lf\psi_\La\psi_\Lb
\vphantom{\left[\ppp\right]}\right]~+\nonumber \\
&&{{\Lf^2+\La^2-\Lb^2\over2r^2}}\left[\vphantom{\left(\ppp\right)}
(\Xi^r-{1\over r}\Psi)_\Lf\psi_\La \PD_r\dep_\Lb
+r\PD_r({1\over r}\Psi)_\Lf\dep_\La\xi^r_\Lb
-r\PD_r({1\over r}\delta P)_\Lf\psi_\La\xi^r_\Lb\right.\nonumber \\
&&-\rho r\PD_r(r^{-1}\delta U)_\Lf\psi_\La\xi^r_\Lb
-\left(\delta P+{1\over r}\pp\Psi\right)_\Lf\psi_\La\divxi_\Lb
-{d\over dr}\left({1\over r}\pp\right)\Psi_\Lf\psi_\La\xi^r_\Lb
\left.\vphantom{\left(\ppp\right)}\right]~+\nonumber \\
&&\left[\vphantom{\left(\ppp\right)}
\PD_r\Xi^r_\Lf\xi_\La\PD_r\dep_\Lb -\left(\PD_r\delta P +\pp\divXi
+\ppp\Xi^r\right)_\Lf\xi^r_\La\divxi_\Lb\right.\nonumber \\
&&\qquad-{\textstyle 1\over2}\left(\delta P +\gamma(\gamma-2)p\divXi
+\pp\Xi^r\right)_\Lf\divxi_\La\divxi_\Lb \nonumber \\
&&\qquad -{\textstyle 1\over2}
\left(\ppp\divXi+{d^3p\over dr^3}\Xi^r
+\rho{d^3U\over dr^3}\Xi^r+\rho\PD_r^2\delta U+\PD_r^2\delta P
\right)_\Lf\xi^r_\La\xi^r_\Lb
\left.\left.\vphantom{\left(\ppp\right)}\right]~~~\right\}
\label{eq:couplings}
\end{eqnarray}
Every term in this expression is formally of order $(\xi^r/H)^2 p$,
where $\xi^r=\xi^r_\La$ or $\xi^r_\Lb$, since
$\psi_\La\sim R\xi^r_\La/\ell$. The 3-mode coupling coefficent,
$\kappa_{\alpha\beta\gamma}$, is obtained by integrating the above
expression over the radius of the star i.e.

$$ \kappa_{\alpha\beta\gamma} = \int dr\,r^2 \int d^2\Omega\,{\cal L}_3,
   \eqno(A18)$$
Note that all the eigenfuctions in equation (A17) are normalized to unit
energy.

When all three interacting modes have wavelengths
$(\lambda_1,\lambda_2,\lambda_3)\ll (R_*,H)$,
then the first term on the righthand side of (\ref{eq:lagxi})
dominates the rest
by a factor $\sim H/\max(\lambda_1,\lambda_2,\lambda_3)$.
This justifies our statement in the text that the coefficient
$\kappa_{\beta\gamma\gamma}$ coupling daughter to granddaughter modes
exceeds the coupling $\cplb$ of the primary mode to its daughters by
a factor $\sim\ell_\beta$.
Physically this can be understood, or at least remembered, by noting
that the parametric growth rate scales as the shear
$\sim\omega\xi/\lambda$ in the parent mode.

\newpage
\section{Rotational splitting of eigenfrequencies}
\setcounter{equation}{0}

The rotational splitting of mode frequencies as seen in the mean
rotational frame of the star is given by (e.g. \cite{Stix:91})

\begin{equation}
 \Delta\nu_{n\ell m}  \equiv \nu_{n\ell m}-\nu_{n\ell 0} =
  -m C_{n\ell}\Omega_0/2\pi,
\end{equation}
where $\Omega_0$ is the mean angular speed of the star,
\begin{equation}
\label{eq:cnleq}
 C_{n\ell} = { \int d^3x\;  \rho \left[ \xi_h^2 + 2\xi_r\xi_h -
   (\Delta\Omega/\Omega_0)\left\{ (\xi_r-\xi_h)^2 + (\ell^2+\ell-2)\xi_h^2
   \right\}\right] \over \int d^3x\;  \rho \left[ \xi_r^2 +
   \ell(\ell+1)\xi_h^2\right] },
\end{equation}
and
\begin{equation}
  \Delta\Omega(r) \equiv \Omega(r) - \Omega_0.
\end{equation}
For high order g-modes $\xi_r \sim i \ell\xi_h$, and thus
$\int d^3x\;\rho\xi_r\xi_h \approx 0$. Substituting these into equation
(\ref{eq:cnleq}) we find that for a rigidly rotating star
\begin{equation}
C_{n\ell} \approx {1\over 2\ell^2},
\end{equation}
and so the frequency difference between modes of $m=\ell$ and $-\ell$ is
$\Omega_0/2\pi\ell$. For a differentially rotating star the splitting
coefficient is
\begin{equation}
C_{n\ell} \approx {1\over 2\ell^2} - {\overline{\Delta\Omega}\over \Omega_0},
\end{equation}
where $\overline{\Delta\Omega}$ is the mean value of differential rotation
in the radiative interior of star where high order g-modes are trapped.
This leads to $\Delta\nu_{n\ell m}\approx -m\overline{\Delta\Omega}/2\pi$
if $\ell\gg 1$.
However, if this formula predicts $\Delta\nu_{n\ell m}>N_B$, the
\BV frequency, then individual g-modes are trapped within regions
smaller than the entire radiative zone, and the spectrum of mode
frequencies becomes continuous.

\newpage
\section{Frequency shifts and saturation}
\setcounter{equation}{0}

Mechanical oscillations and waves are generally anharmonic: their
natural frequency depends upon amplitude.
Unless other processes intervene earlier, a
parametric instability will saturate at an amplitude such
that the resonance condition (\ref{eq:instrip}) is no longer satisfied
because of the anharmonic frequency shift.

For example, a generic one-dimensional oscillator with Hamiltonian
\begin{equation}
H= \frac{1}{2m}p^2 + m\omega_0^2 q^2\left\{\frac{1}{2}
+\frac{\kappa}{3a_0} q
+\frac{\mu}{4a_0^2}q^2 + O[(q/a_0)^3]\right\}
\label{eq:singleH}
\end{equation}
has free oscillations (cf. \cite{LL})
\begin{eqnarray}
q(t) &=& a\cos[\omega(a) t], \nonumber \\
\omega(a) &=& \omega_0\left[1 +\left(\frac{3}{8}\mu-
\frac{5}{12}\kappa^2\right)\left(
\frac{a}{a_0}\right)^2 + O\left(\frac{a^4}{a_0^4}\right)\right].
\end{eqnarray}
Here $a_0$, a characteristic amplitude at which anharmonicity becomes
important, is chosen
so that the larger of $\mu$ and $\kappa^2$ is of order unity.
Suppose this oscillator is coupled to another in
such a way that very small oscillations are parametrically unstable with
growth rate $\eta=\epsilon\omega_0$.
The range of frequencies over which parametric instability occurs is
$\Delta\omega\approx\eta$.
If (for simplicity) parametric resonance is exact at infinitesimal
amplitude, then growth stops when $|\omega(a)-\omega_0|\approx\eta$,
hence $a\sim\epsilon^{1/2} a_0$.

Waves in fluids are normally strongly nonlinear when $ka\sim1$,
where $a=\max(|\vec\xi|)$ is the peak displacement and $k=2\pi/\lambda$
is the wavenumber.
That is, we may normally take $a_0=\lambda/2\pi$.
This is because the eulerian dynamical equations contain the advective
derivative $\partial_t +\vec v\cdot\vec\nabla$, and for small perturbations
of a static equilibrium, the ratio of the second (nonlinear) part of this
operator to the first is $O(k\delta v/\omega)=O(ka)$.
A single g mode, however, is an exception to this general rule if
the wavelength is small compared to a pressure or density scale height,
$H$.
Such a wave is nearly incompressible, so that if $\delta\vec v\propto
\exp(i\vec k\cdot\vec x)$ in a local plane-wave approximation, then
$\vec\nabla\cdot\delta\vec v$ is $\sim \delta v/H$
instead of $\sim k\delta v$.
It follows that the operator $\vec v\cdot\vec\nabla=O(\delta v/H)$ when
applied to any fluid perturbation whose WKBJ wavenumber is parallel
to that of $\delta\vec v$.
One can check that other nonlinear terms vanish to leading order in
$(kH)^{-1}$.
For example, the dynamical equation for the vorticity
$\vec\nabla\times\vec v$ contains the source term
$\rho^{-2}(\vec\nabla\rho\times\vec\nabla p)$; but for a single
wave, the gradients of the density and pressure perturbations are both
approximately parallel to the wavenumber, so that
$\vec\nabla\delta\rho\times\vec\nabla\delta p$ vanishes to leading
order in $(kH)^{-1}$.

We note in passing that nonlinear interactions involving two or
more nonparallel wavenumbers need not be similarly suppressed.
In parametric instability, for example, the primary and daughter
modes have nonparallel wavenumbers.
Anharmonicity is a special nonlinearity because it involves
self-interaction of a single mode.

Consequently, the anharmonic frequency shift of a WKB g mode is
\begin{equation}
\omega(a)-\omega_0= O\left(\omega_0\frac{a^2}{H^2}\right).
\end{equation}
Notice that this does not depend on mode wavelength.
Equating this shift to the width $\Delta\omega\sim\epsilon\omega_0$
of the parametric resonance caused by a fundamental mode of
dimensionless amplitude $\epsilon=\delta R_*/R_*$, one predicts
saturation when the small-scale mode has amplitude
$a\sim\epsilon^{1/2}H$.
In the radiative core, however, $H\sim R$, so the small-scale mode
would have more energy than the large-scale one by a factor of order
$(a/\delta R_*)^2\sim\epsilon^{-1}$.

Clearly, other nonlinear processes must intervene to halt the growth of
the small-scale mode before frequency shifts become important.
In the situation contemplated here,  we foresee only two such
processes: one is the decrease in the parametric growth rate as the
energy in the primary mode declines; and  the other is the destabilization
of granddaughter modes.
Both processes have been discussed in \S 2.2.

\newpage
\begin{center}
{\bf Table 1}\\
Mode coupling coefficients for quadrupole f- and g-modes of a
solar model with high degree g-modes of half the frequency.
First two columns are the number of radial nodes and the frequency
of the primary mode ($\alpha$).
Third column is coupling coefficent for parametric decay into identical
daughter modes ($\beta$).
Last column gives the fractional surface displacement amplitude of
the primary mode when its energy is 10$^{42}$ erg.
\end{center}
\begin{center}
\begin{tabular}{|c|c|c|c|} \hline
$n_\alpha$ (order) & $\nu_\alpha$
& $\kappa_{\alpha\beta\beta}$ & $\delta$R/R$_*$ when \\
& ~~~($\mu$Hz)~~~  & (erg$^{-1/2}$)  & E$_\alpha$ = 10$^{42}$ erg\\ \hline
            &            &               &                           \\
     0 &   372.2 &   ~4.19$\times 10^{-25}$  &  1.7$\times 10^{-3}$ \\
     1 &   300.8 &  -4.43$\times 10^{-25}$ &  1.8$\times 10^{-3}$ \\
     2 &   262.7 &  ~1.13$\times 10^{-24}$ &  1.6$\times 10^{-3}$ \\
     3 &   225.7 &  -1.31$\times 10^{-24}$ &  1.1$\times 10^{-3}$ \\
     4 &   195.6 &   ~2.13$\times 10^{-24}$ &  8.7$\times 10^{-4}$ \\
     5 &   170.9 &  -2.86$\times 10^{-24}$ &  7.2$\times 10^{-4}$ \\
     6 &   151.1 &  ~4.50$\times 10^{-24}$ &  6.3$\times 10^{-4}$ \\
     7 &   135.0 &  -7.00$\times 10^{-24}$ &  6.0$\times 10^{-4}$ \\
     8 &   121.9 &   ~7.86$\times 10^{-24}$ &  5.9$\times 10^{-4}$ \\
     9 &   111.0 &  -8.61$\times 10^{-24}$ &  5.7$\times 10^{-4}$ \\
    10 &   101.7 &   ~1.48$\times 10^{-23}$ &  5.8$\times 10^{-4}$ \\
    11 &    93.9 &  -1.49$\times 10^{-23}$ &  5.9$\times 10^{-4}$ \\
    12 &    87.1 &   ~2.17$\times 10^{-23}$ &  6.0$\times 10^{-4}$ \\ \hline
\end{tabular}
\end{center}

\newpage
\begin{center}
{\bf Table 2}\\
The size of convection zone in main sequence stars of metalicity ($z$) equal
to 0.0001 and age 13.0 Gyr.
The first column is the mass of the star in the unit of solar mass,
the second column is the effective
temperature, the third column is the thickness of the convection zone
divided by the radius of the star, and the last column is the fractional
mass in the convection zone. These numbers, obtained using the
Yale stellar evolution code, were kindly provided by Constantine Deliyannis.

\end{center}
\begin{center}
\begin{tabular}{|c|c|c|c|} \hline
M$_*$ & T$_{\rm eff}$
& R$_c$/R$_*$ & M$_c$/M$_*$ \\
\hline
     0.50 &   4567 &   0.345 &  0.1940 \\
     0.65 &  5509 &   0.230 &  0.0274 \\
     0.75 &  6227 &  0.135 &  0.0012 \\ \hline
\end{tabular}
\end{center}

\newpage
\centerline{\bf Figure Captions}
\bigskip

\noindent FIG. 1.--- Parametric nonlinear interaction involving $10^3$ modes.
The amplitude of the primary mode is shown as a function of time.

\medskip
\noindent FIG 2.--- The upper panel shows the radial displacement
eigenfunction of the f-mode of energy = 10$^{45}$ erg (solid line),
and the superposition of radial p-mode displacement-functions generated as a
result of nonlinear couplings to the f-mode (dash-dot curve). The
lower panel shows the radial derivative of wave functions shown in the
upper panel.

\medskip
\noindent FIG. 3.--- Energy input per orbit (in erg) into quadrupole f- and
a few low order g-modes of $m=\ell=2$, as a function of periastron distance.
The mass of the star is $0.7~M_{\sun}$, and the orbital eccentricity is 1.0.
The star has been modelled as a polytrope of index 3 with the ratio of
specific heat taken to be 5/3.
{}From top to bottom at left ($d_{\rm peri}/R_*=2.5$) the modes are
$g_2,~f,~g_4,~g_6,~g_8$, and $g_{10}$.

\medskip
\noindent FIG 4.--- The log of nonlinear damping time (in days) for a few low
order quadrupole modes of a $0.7~M_{\sun}$ radiative star as a function
of periastron distance. The linear radiative damping time for all of
these modes are about $10^4$ years.
Curves correspond to same modes as in Fig.~3: that is, from top to bottom
at $d_{\rm peri}/R_*=5$, $f,~g_2,~g_4,~g_6,~g_8$, and $g_{10}$.

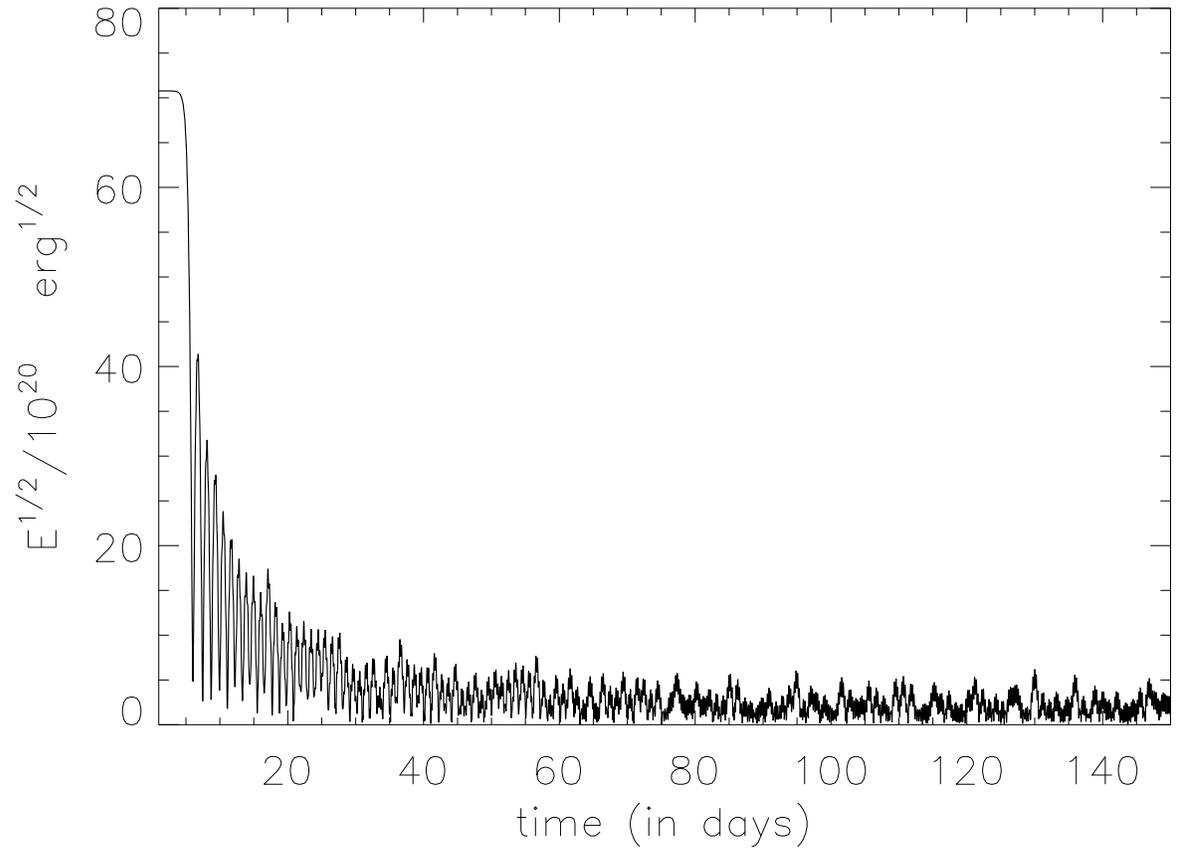
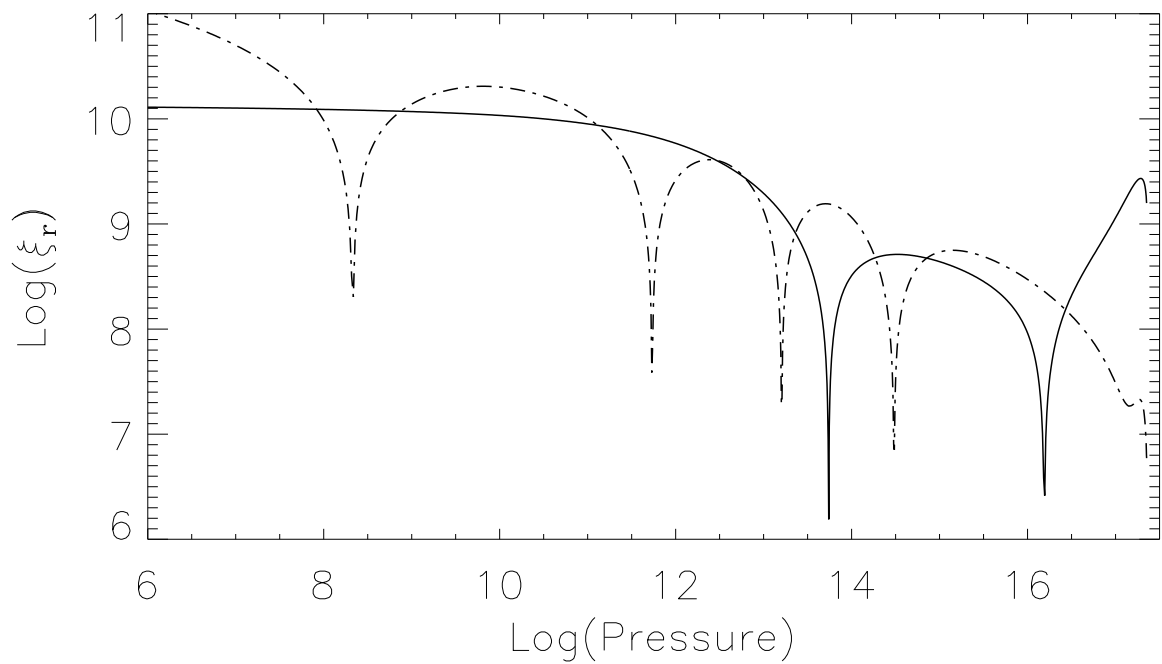
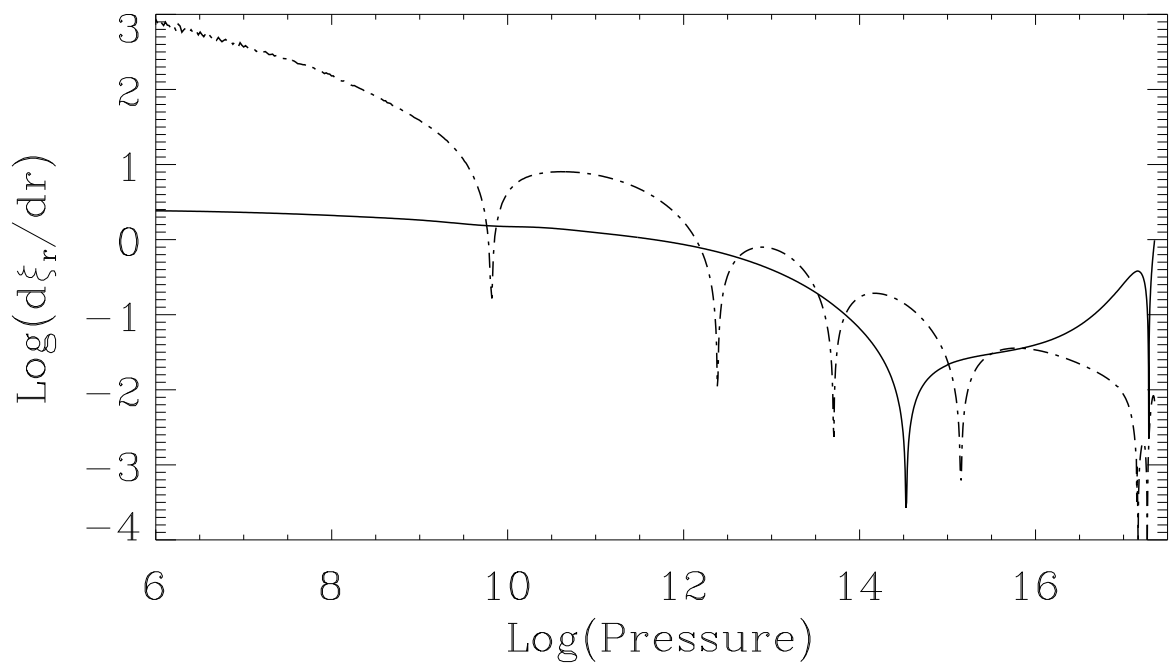
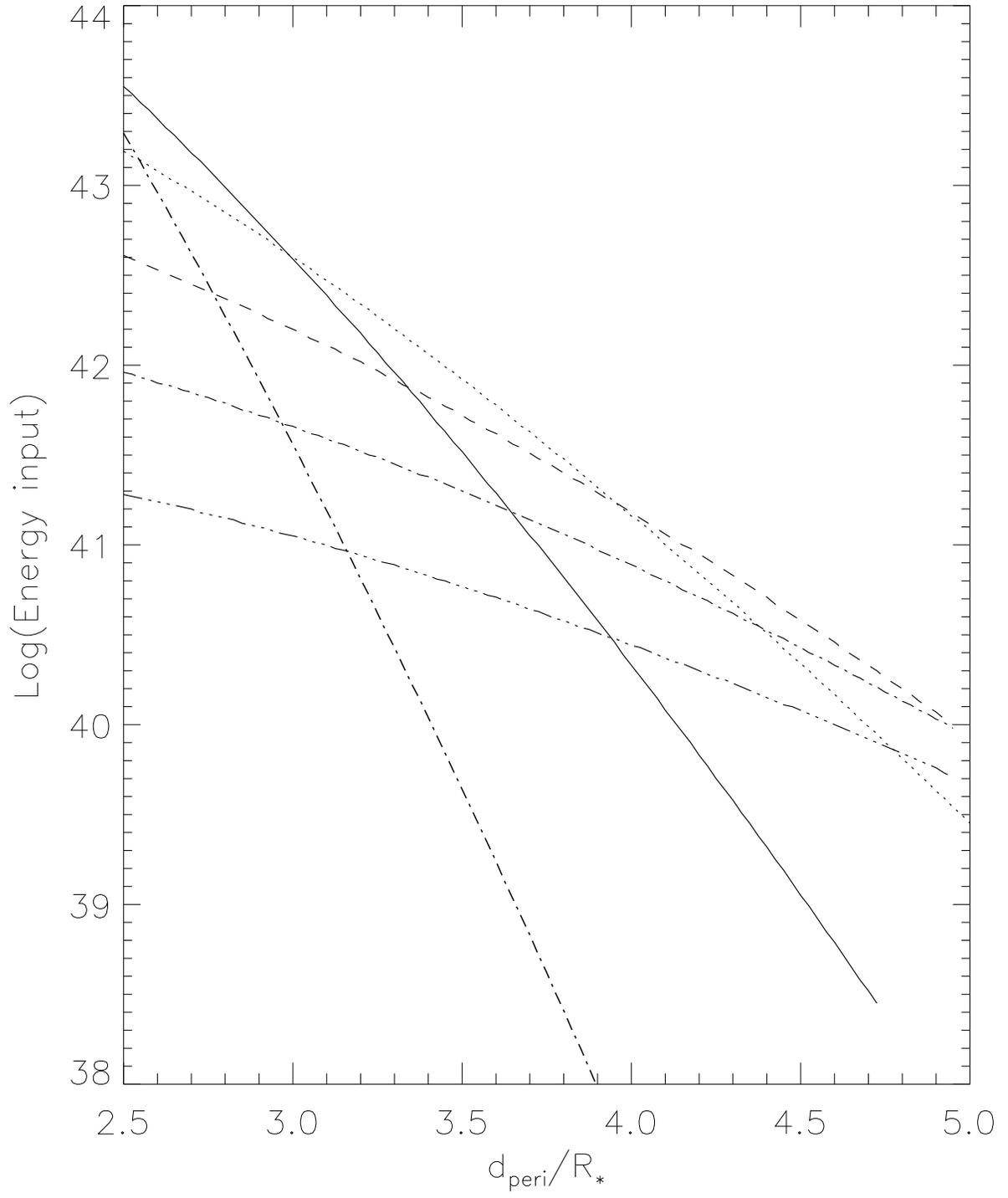
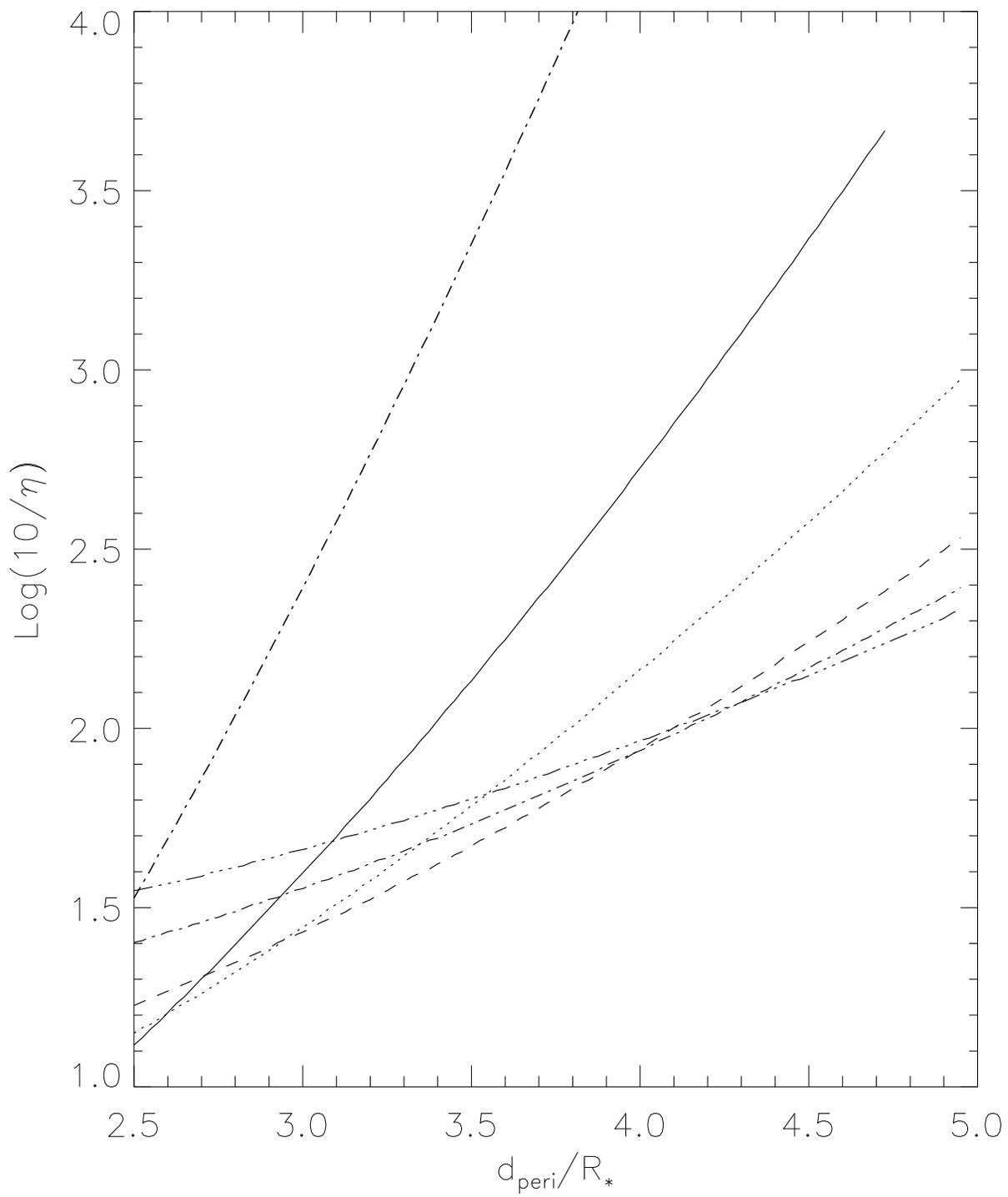
\end{document}